\documentclass[aps,pra,reprint,amsmath,amssymb,superscriptaddress,longbibliography]{revtex4-2}

\usepackage{graphicx}
\usepackage{xcolor}
\usepackage{bm}
\usepackage{siunitx}
\usepackage{epsfig}
\usepackage{dcolumn}
\usepackage{epstopdf}
\usepackage{multirow}
\usepackage{booktabs}
\usepackage{gensymb}
\usepackage{kantlipsum}
\usepackage[normalem]{ulem}
\usepackage{braket}
\usepackage{physics}
\usepackage{url}
\usepackage[utf8]{inputenc}
\usepackage[T1]{fontenc}
\usepackage{mathptmx}
\usepackage{lineno}
\usepackage{comment}
\usepackage{soul}
\usepackage{float}
\usepackage{hyperref}

\begin{document}

\title{High precision micro-optical elements on fiber facets via focused-ion beam machining} 

\author{Raman Kumar}
\affiliation{Quantum Information Science and Technology, Brookhaven National Laboratory, Upton, NY 11973, USA}

\author{Sebastian Will}
\affiliation{Quantum Information Science and Technology, Brookhaven National Laboratory, Upton, NY 11973, USA}
\affiliation{Department of Physics, Columbia University, New York, NY 10027, USA}

\date{\today}

\begin{abstract}
Fiber-integrated micro-optical elements promise a scalable approach to photon collection and beam shaping for quantum information processing. Here, we demonstrate single-step micro-machining of micro-spherical, micro-spiral, and micro-axicon structures directly onto the core of single-mode optical fibers using focused-ion beam (FIB) machining with nanometer-scale precision. Atomic force microscopy reveals that micro-concave and micro-convex spherical surfaces achieve shape accuracies of $\lambda/50$ and $\lambda/33$ (at $\lambda = 633$~nm), respectively. Via optical characterization using a He-Ne laser (633~nm),  we verify for the micro-spiral and -axicon structures the expected far-field donut beam patterns. Via Mach-Zehnder interferometry, we verify the azimuthal and radial phase structure of the light emitted from the spiral and axicon fibers, respectively. Surface metrology further shows that FIB processing introduces only a small increase in nanoscale roughness while preserving sub-nanometer RMS surface quality. The monolithically integrated fiber micro-optic elements demonstrated here enable broad use in quantum technology, for example, fiber micro-cavities for cavity quantum electrodynamics, beam shaping for neutral atom trapping, and the generation of structured light for free-space quantum network links.
\end{abstract}

\maketitle

\section{Introduction}

Micro-optical elements fabricated on fiber facets have found applications in a broad range of fields including endoscopic optical coherence tomography~\cite{Lorenser2012} and integrated photonics~\cite{Cabrini2006}. More recently, optical elements based on microstructured fiber facets have gained increased attention for quantum applications, in particular to enhance the interaction between atoms and photons in optical resonator structures~\cite{Grinkemeyer2025}. This is of particular interest for quantum computing platforms based on neutral atom arrays~\cite{Bluvstein2024} and trapped ions~\cite{Bernardini2024}. Using a photonic bus, quantum computing modules can be linked to effectively scale system sizes. In addition, the enhancement of the readout speed in atom-based quantum computing platforms is of great interest.

Optical cavities provide a promising route to address these  challenges by enhancing atom-photon interactions~\cite{Reiserer2022}. Experimental implementations using macroscopic cavities in the telecom band have demonstrated coherent atom-cavity coupling~\cite{Covey2023,Li2024,LiL2025} and remote entanglement distribution over distances exceeding 100 km~\cite{Lu2026}. While these results are encouraging, macroscopic resonators come with relatively large mode volumes and limited cooperativity. Macroscopic optical elements also complicate scalability which is crucial for practical quantum technologies.  Microcavity approaches, by contrast, enable dramatically reduced mode volumes and enhanced light-matter coupling strengths, making them particularly attractive for accessing the strong-coupling regime~\cite{Sinclair2025}. Fiber-integrated geometries are of particular interest. They allow direct coupling of cavity photons into guided fiber modes, facilitating remote entanglement generation and distribution.  Integrated fiber-based microcavity systems have been shown to realize high cooperativity and high fidelity two-qubit quantum gate operations~\cite{Gehr2010,Grinkemeyer2025}. Similarly, fiber micro-cavities have also been used to demonstrate strong coupling in trapped ion systems~\cite{Takahashi2020}. 

Fiber micro-cavities require the fabrication of micro-concave structures on top of fiber facets. Pioneering work has realized such structures using CO$_2$ laser ablation~\cite{Hunger2010}. Subsequent work has combined focused-ion beam (FIB) processing and laser reflow for surface smoothing~\cite{Maier2025}, as well as alternative microfabrication and etching approaches~\cite{Ding2026,kumar2021orb}. While these techniques can produce high-quality structures, they offer limited flexibility in defining more general surface profiles. In contrast, FIB milling promises to provide direct, three-dimensional, free-form fabrication with nanometer-scale resolution which can open new opportunities beyond spherically symmetric micro-structures. For example, a fiber-integrated convex lens could enable in situ focusing for cold-atom experiments~\cite{Glicenstein2021}, while an axicon on a fiber tip allows direct generation of Bessel beams that can offer resilience to turbulence in free-space quantum links~\cite{McLaren2014,Li2017}. A spiral element on a fiber tip can impart orbital angular momentum on photons and enable higher-order entanglement in practical quantum networking~\cite{Krenn2017}. Similarly, tailored fiber micro-optics may enhance coupling efficiency in astronomical instrumentation~\cite{Jovanovic2017}. The capability of FIB machining to directly produce surfaces meeting the stringent requirements of high-precision applications remains to be established.

In this work, we demonstrate a single-step micro-machining process, together with comprehensive structural and optical characterization, of micro-optical elements fabricated directly on single-mode fiber cores using focused ion beam (FIB) milling. Specifically, we demonstrate micro-concave, micro-convex, micro-spiral, and micro-axicon structures. To enable accurate centering of the structures to the guided mode, we develop a process that exposes the fiber core using a buffered oxide etch step and image it using in-situ electron microscopy. We characterize the structural quality and their optical performance via scanning electron microscopy (SEM), atomic force microscopy (AFM), focal-field measurements, far-field imaging, and Mach-Zehnder interferometry. Our results show that FIB milling can reproducibly achieve shape accuracies better than $\lambda/33$, establishing this technique as a route toward fiber-integrated precision micro-optics.

\begin{figure*}
\includegraphics[width=0.66\textwidth]{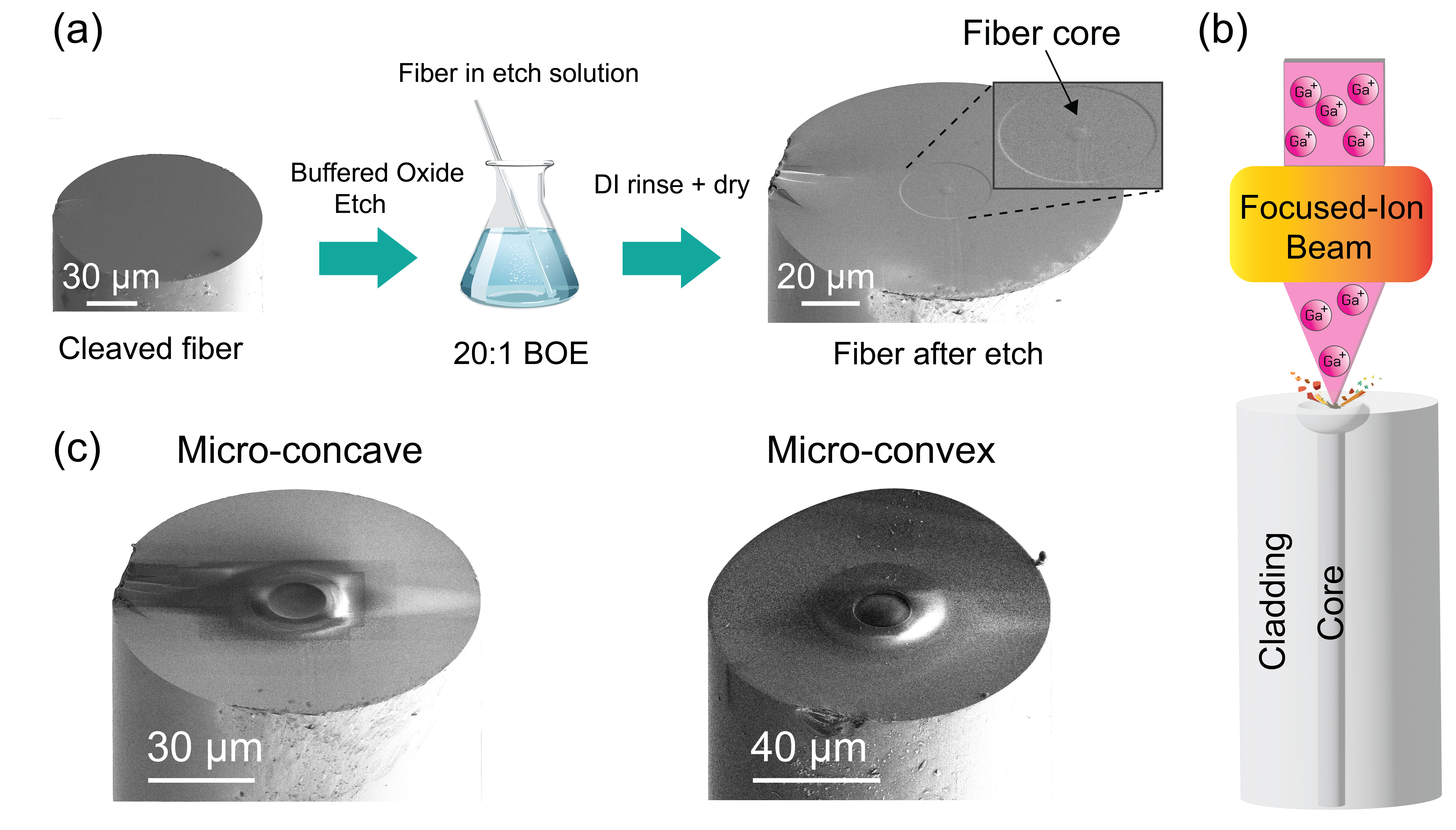}
\caption{Fabrication of micro-optical elements directly on a single-mode fiber facet. \textbf{(a)}~Selective chemical etching of optical fibers in 20:1 buffered oxide etch (BOE) for 15~min reveals three distinct etch-contrast regions: the central core, an intermediate annular region, and the outer cladding, enabling deterministic alignment. Zoomed-in inset on the after etch image shows fiber core localization. \textbf{(b)}~Schematic illustration of FIB processing on the core of the fiber with precise alignment. \textbf{(c,d)}~Scanning electron microscope (SEM) images of representative micro-concave and micro-convex microstructures fabricated on-axis. The structures are centered on the fiber core with sub-micron scale placement accuracy, enabling direct integration with guided modes.}
\label{fig:fabrication}
\end{figure*}

\section{Fabrication}

\subsection{Fiber Preparation and Core Identification}

We use commercially available single-mode fibers with a nominal mode-field diameter of $\sim$\SI{4}{\micro\meter} at 633~nm. A short piece of fiber ($\sim$4~inches in length) is cut from the spool and the end facets are cleaved using a Fujikura CT-50 cleaver. The fiber piece is cleaned using isopropyl alcohol. An SEM image of a typical cleaved and cleaned fiber end is shown in Fig.~\ref{fig:fabrication}(a). 

It is of critical importance to precisely center the micro-optical structures on the guided mode of the fiber. To this end the fiber core needs to be located precisely prior to fabrication.  We achieve this by etching the fiber tip with a hydrofluoric acid-based solution~\cite{Janeiro2016} and exploiting the differential etch rate of silica with different doping levels. The freshly cleaved fiber is immersed in a 20:1 buffered oxide etch (BOE) solution for 15~minutes [Fig.~\ref{fig:fabrication}(a)], followed by rinsing in deionized water and drying. Following the etching, the fiber facet reveals three distinguishable regions rather than a simple core-cladding structure: a central core, a surrounding annular intermediate region, and the outer cladding. The observed ring-like intermediate contrast is consistent with a differentially doped region exhibiting a different etch rate from both the core and the surrounding silica. The central core in the form of a raised pedestal with a typical height of $\sim$\SI{100}{\nano\meter} is readily identifiable using SEM as shown in Fig.~\ref{fig:fabrication}(a), with a zoomed-in inset added for more clarity.

The ability to precisely identify the fiber core addresses a key limitation of prior approaches~\cite{Maier2025,Ding2026}, where insufficient core localization might lead to misalignment of the fabricated structure with respect to the guided mode, increased insertion loss, and limited optical performance. Finally, we coat the fibers with a thin ($\sim$\SI{10}{\nano\meter}) layer of gold which is deposited using an ion sputter coater (MCM-100). Due to the thin gold layer, the fiber facet becomes electrically conductive, avoiding charging effects during FIB processing and SEM imaging.

\subsection{Focused-Ion Beam Processing}

FIB milling is performed using a Thermo Fisher Helios G5 dual-beam instrument equipped with a liquid gallium metal ion source. The fiber is mounted in a custom holder designed to position the cleaved facet normal to the ion beam. More details about fiber mounting and feature design are given in Appendix~\ref{app:mounting}. It is critical to ensure normal incidence of the ion beam on the fiber facet for accurate fabrication results. In the Helios G5 dual-beam geometry, the electron and ion columns are oriented at a relative angle of $52^\circ$ as shown in Fig.~\ref{fig:appendixA}(b), with the two beams converging at the eucentric height of the sample stage~\cite{Holdford2005}. To achieve normal incidence of the ion beam on the fiber facet, the sample stage is first brought to the eucentric height by iteratively adjusting the stage $z$-position while tilting until the feature of interest remains stationary in the electron beam image. The stage is then tilted to $52^\circ$ so that the fiber facet is perpendicular to the ion column axis. This alignment procedure is performed before each session and is critical for achieving rotationally symmetric profiles and reproducible phase geometries across different fibers. The electron column (5~kV, 25~pA) is subsequently used first for high-resolution imaging of the etched core, after which the fiber is aligned such that the FIB pattern is centered on the core. A schematic of the FIB process is shown in Fig.~\ref{fig:fabrication}(b).

\begin{figure*}
\includegraphics[width=0.66\textwidth]{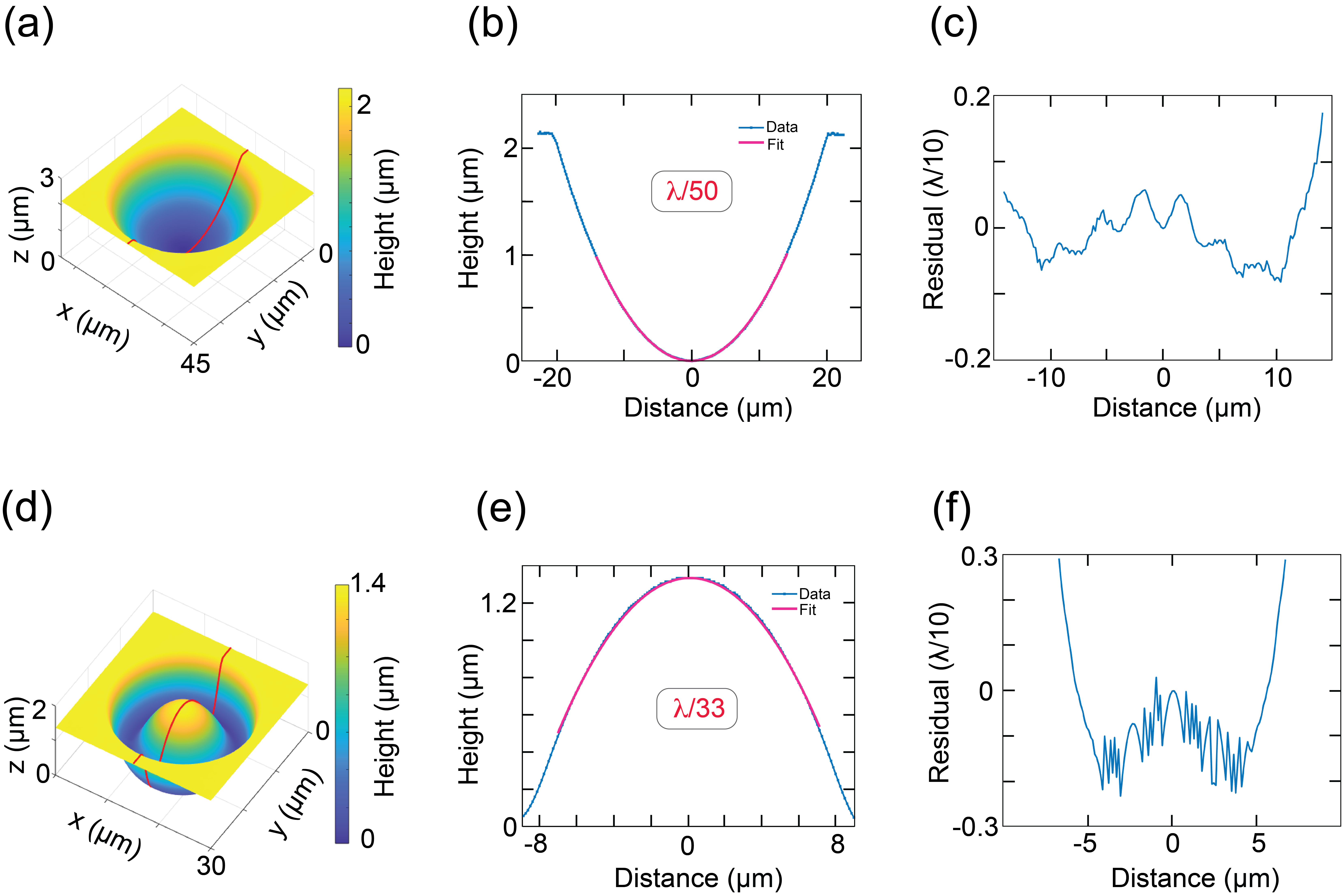}
\caption{Surface metrology of FIB-fabricated spherical micro-optics. \textbf{(a,d)}~Three-dimensional atomic force microscope (AFM) topography images for micro-concave (scan area: $45~\si{\micro\meter} \times 45~\si{\micro\meter}$) and micro-convex (scan area: $30~\si{\micro\meter} \times 30~\si{\micro\meter}$) features, respectively. The red line cut across the center is used for curve fitting analysis. \textbf{(b,e)}~Line cut data across the diameter and a spherical curve fit demonstrating $\lambda/50$ and $\lambda/33$ residuals over 50\% of the total area for both micro-concave and micro-convex surfaces, respectively. The curve-fitted ROC of micro-concave and micro-convex surfaces are \SI{100.76}{\micro\meter} and \SI{28.49}{\micro\meter}, respectively. \textbf{(c,f)}~Plots of resulting residuals after curve-fitting for micro-concave and micro-convex surfaces, respectively, in $\lambda/10$ units at $\lambda = 633$~nm.}
\label{fig:afm}
\end{figure*}

Micro-optical elements are defined using bitmap pattern files ($512 \times 512$ pixels) in which the grayscale value encodes the local milling depth [Fig.~\ref{fig:appendixA}(c) in Appendix~\ref{app:mounting}]. The grayscale pixel values are associated with different dwell times upon FIB exposure. For spherical elements (concave and convex), the pattern is generated from the analytical height profile of a sphere with the desired radius of curvature. For the micro-spiral phase plate, the pattern encodes a linear azimuthal height variation producing a $2\pi$ phase ramp around the optical axis at a wavelength of 633 nm. For the micro-axicon, a conical height profile is used. Typical fabrication parameters include a beam voltage of 30~kV, a beam current of 0.75~nA for micro-spiral/axicon and 2.6~nA for micro-concave/convex, and a dwell time ($1$-$10~\si{\micro\second}$) optimized for the target depth. The milled area is defined by the bitmap dimensions, with a typical element diameter of \SI{10}{\micro\meter} for the phase elements and larger areas (up to \SI{40}{\micro\meter}) for the spherical elements. Following FIB fabrication, the sacrificial gold layer was removed using aqua regia (HCl:HNO$_3$ = 3:1). Aqua regia readily dissolves Au but does not attack or etch silica, ensuring that the underlying fiber surface remains unaffected during the metal-removal step. Typical SEM images of a micro-concave (diameter: \SI{20}{\micro\meter}) and a micro-convex (diameter: \SI{20}{\micro\meter}) feature are shown in Figs.~\ref{fig:fabrication}(c) and~\ref{fig:fabrication}(d), respectively. Furthermore, the FIB process adds negligible surface roughness in comparison to optical wavelengths as discussed in Appendix~\ref{app:roughness}.

A key advantage of the single-step FIB micro-machining approach presented here is the ability to precisely fabricate concave, convex, and other structures, solely via material removal through grayscale milling. This flexibility enables a wide range of optical functionalities from a single fabrication platform without the need for material deposition or template transfer~\cite{Principe2017}.


\section{Characterization}

\subsection{Micro-spherical elements}

The surface topography of the FIB-fabricated micro-spherical elements is characterized by atomic force microscopy (AFM; Park Systems FX40). Figures~\ref{fig:afm}(a) and~\ref{fig:afm}(d) present three-dimensional AFM topography maps for a micro-concave surface and a micro-convex surface, respectively. Line cuts through the center of each feature along the $y$-direction are extracted and fitted to a spherical model, yielding radii of curvature (ROC) of \SI{100.76}{\micro\meter} (Design = \SI{100}{\micro\meter}) and \SI{28.49}{\micro\meter} (Design = \SI{30}{\micro\meter}) for the micro-concave and micro-convex surfaces, respectively [Figs.~\ref{fig:afm}(b) and~\ref{fig:afm}(e)]. Corresponding line cuts along the orthogonal $x$-direction yield ROC values of \SI{99.88}{\micro\meter} and \SI{29.15}{\micro\meter}, demonstrating a cross-directional ROC agreement of better than 0.88\% and 2.3\%, respectively. This level of rotational symmetry provides direct evidence of the high degree of sphericity achieved by FIB milling and, to our knowledge, has not been explicitly quantified in previous reports of fiber micro-optical fabrication~\cite{Hunger2010,Maier2025,Ding2026}.

\begin{figure}   
\includegraphics[width=0.75\columnwidth]{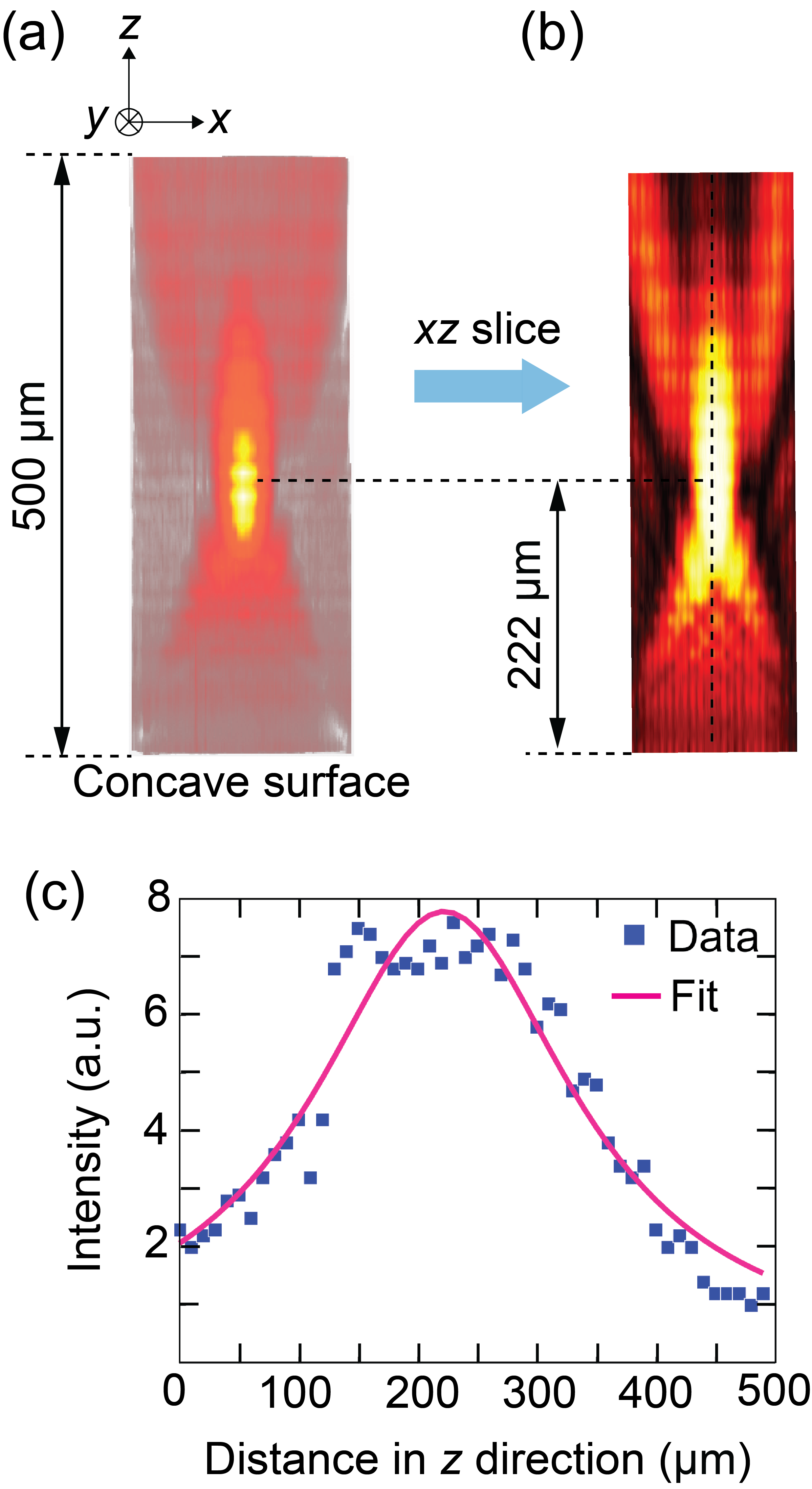}
\caption{Experimental verification of focusing from a micro-concave element. \textbf{(a)}~ Three-dimensional intensity reconstruction obtained from a 50-plane $z$-stack with a step size of \SI{10}{\micro\meter} using a custom-built Helium-Neon laser (633~nm) based microscope. \textbf{(b)}~$xz$ plane cross-section from the 3D volume rendering in (a) at constant $y$. \textbf{(c)}~Data points show a cut across the dashed line of the $xz$ plane image in (b) demonstrating a focal length of about \SI{222}{\micro\meter}. The red line represents the fit to the measured axial peak-intensity evolution using the on-axis Gaussian beam propagation model, $I(z)=I_{\mathrm{bg}}+I_0/[1+((z-z_0)/z_R)^2]$, yielding an effective focal position of $z_0 = 222.2 \pm 3.6~\mu\mathrm{m}$. }
\label{fig:focusing}
\end{figure}

The shape accuracy of the fabricated surfaces is assessed by computing the residuals between the measured height profile and the best-fit spherical surface. For the micro-concave element, the residuals are $\lambda/50$ (at $\lambda = 633$~nm) over the central 50\% of the total feature area [Fig.~\ref{fig:afm}(c)]. For the micro-convex element, the residuals are $\lambda/33$ over the central 50\% of the feature area [Fig.~\ref{fig:afm}(f)]. In standard optical metrology practice, surface quality is evaluated over the optically used clear aperture rather than the full physical extent of the optic~\cite{Smith2007}. For fiber-integrated micro-optics, the relevant aperture is defined by the Gaussian mode of the single-mode fiber: the fiber used here has a mode-field diameter (MFD) of approximately \SI{4}{\micro\meter} at 633~nm. In the worst-case scenario, the propagating optical mode illuminates the central $\sim$\SI{10}{\micro\meter} diameter of the micro-spherical surface, corresponding to roughly 6\% and 25\% of the total fabricated area, respectively. The 50\% fitting regions therefore conservatively encompass the full optically active area with significant margin. The reported residual errors of $\lambda/33$ - $\lambda/50$ should thus be regarded as lower-bound estimates of the achievable surface accuracy. Restricting the fit to a smaller, mode-weighted region yields correspondingly smaller residuals, indicating even higher surface quality within the area most relevant to optical operation.

To place these surface residuals in the context of quantitative optical performance metrics, we compute the equivalent Strehl ratio using the Mar\'{e}chal approximation~\cite{Born1999,Mahajan1983},
\begin{equation}
S \approx \exp\!\left[-(2\pi\sigma)^2\right],
\label{eq:strehl}
\end{equation}
where $\sigma$ is the rms wavefront error expressed in units of the wavelength. The relationship between surface error $\delta$ and wavefront error depends on whether the element is used in reflection or transmission. For a reflective surface, as relevant to fiber Fabry-P\'{e}rot microcavities where the micro-concave element serves as a cavity mirror, the wavefront error is twice the surface error, $\Delta W = 2\delta$, because the light traverses the surface deformation on both the incident and reflected paths. For a transmissive surface at a glass-air interface, the wavefront error is reduced to $\Delta W = (n-1)\delta$, where $n$ is the refractive index of the substrate; for fused silica at 633~nm, $n \approx 1.46$, giving a factor of 0.46 which is approximately four times smaller than the reflective case.

For the micro-concave surface with an rms surface residual of $\lambda/50$ at 633~nm, operation in reflection gives a wavefront error of $\sigma = 2/50 = 0.04$, yielding a Strehl ratio of $S = \exp[-(2\pi \times 0.04)^2] \approx 0.94$. For the micro-convex surface operating in transmission, the surface residual $\lambda/33$ produces $\sigma = 0.46/33 \approx 0.014$, giving $S > 0.99$. Both values far exceed the Mar\'{e}chal criterion ($S \geq 0.80$) that defines diffraction-limited performance~\cite{Born1999,Mahajan1983}, confirming that the FIB-fabricated surfaces are of exceptional optical quality.

To assess the reproducibility of the FIB fabrication process beyond the representative structures shown in Fig.~\ref{fig:afm}, we fabricated and characterized multiple nominally identical micro-spherical elements. For each structure, orthogonal line profiles through the center of the AFM topography were independently fitted to spherical surfaces along the $x$- and $y$-directions. For the micro-concave structures, the fitting region was defined using a maximum residual criterion of $\lambda/50$ at $\lambda=633$~nm. Table~\ref{tab:concave_reproducibility} provides the results for five independently fabricated micro-concave structures. The device-averaged radii of curvature span $98.82$ to  $100.32~\si{\micro\meter}$, yielding an ensemble average of $99.66\pm0.56~\si{\micro\meter}$ (mean $\pm$ standard deviation, $n=5$). The corresponding coefficient of variation is as low as $0.56\%$, demonstrating a high degree of device-to-device reproducibility for nominally $\sim100~\si{\micro\meter}$-ROC structures. In addition, the independently fitted $x$- and $y$-direction ROCs agree to within $0.9$ to $3.9\%$ for all five structures, with an average cross-directional difference of approximately $2.5\%$. These measurements demonstrate that the FIB process reproducibly reaches both the targeted curvature and a high degree of rotational symmetry across independently fabricated devices.

\begin{table}[t]
\centering
\caption{Reproducibility of independently fabricated micro-concave structures. $\mathrm{ROC}_x$ and $\mathrm{ROC}_y$ are obtained from independent spherical fits to orthogonal AFM line profiles. The fitting region for each structure is defined using a residual criterion of $\lambda/50$ at $\lambda=633$~nm. The mean ROC denotes $(\mathrm{ROC}_x+\mathrm{ROC}_y)/2$.}
\label{tab:concave_reproducibility}
\begin{tabular}{ccccc}
\toprule
Structure &
$\mathrm{ROC}_x$ &
$\mathrm{ROC}_y$ &
Mean ROC &
$x$--$y$ difference \\
&
($\si{\micro\meter}$) &
($\si{\micro\meter}$) &
($\si{\micro\meter}$) &
(\%) \\
\midrule
1 & 100.11 & 97.52 & 98.82  & 2.62 \\
2 & 101.19 & 98.65 & 99.92  & 2.54 \\
3 & 100.76 & 99.88 & 100.32 & 0.88 \\
4 & 101.46 & 97.55 & 99.51  & 3.93 \\
5 & 101.09 & 98.44 & 99.77  & 2.66 \\
\midrule
\multicolumn{3}{l}{Mean $\pm$ SD} &
$99.66\pm0.56$ &
$2.53\pm1.09$ \\
\bottomrule
\end{tabular}
\end{table}

A similar analysis was performed for three independently fabricated micro-convex structures using a residual criterion of $\lambda/33$ at 633~nm. The resulting device-averaged radii of curvature were $28.82$, $27.44$, and $27.75~\si{\micro\meter}$, corresponding to an ensemble average of $28.00\pm0.72~\si{\micro\meter}$. The orthogonal $x$- and $y$-direction fits differed by only $2.3$ to $5.0\%$, demonstrating that the reproducibility observed for the concave structures extends to the more involved micro-convex geometries as well.

The optical focusing performance of the micro-concave element is verified by $z$-stack imaging. For this measurement, a representative micro-concave structure (diameter: \SI{45}{\micro\meter}; ROC: \SI{106.13}{\micro\meter}) is fabricated on a high-purity fused silica substrate (Corning 7980) to enable transmission-mode characterization. The spherical elements demonstrated here are not designed to mode-match the native guided mode of the single-mode fiber into free space. Because the guided mode samples only the nearly planar central portion of these relatively large-radius spherical surfaces, their full focusing response is more appropriately characterized using a substrate-based structure under free-space illumination. A He-Ne laser (633~nm) illuminates the concave surface, and a microscope objective captures the transmitted intensity distribution at 50 axial positions with a step size of \SI{10}{\micro\meter}, from which a three-dimensional focal volume is reconstructed as shown in Fig.~\ref{fig:focusing}(a). A cross-sectional ($xz$) slice through the volume reveals a clearly defined focal region [Fig.~\ref{fig:focusing}(b)], and an intensity line cut along the propagation axis is shown in Fig.~\ref{fig:focusing}(c). The axial evolution of the peak intensity was fitted using the on-axis Gaussian beam propagation model, $I(z)=I_{\mathrm{bg}}+I_0/[1+((z-z_0)/z_R)^2]$, yielding a focal position of $z_0 = 222.2 \pm 3.6~\mu\mathrm{m}$, where the quoted uncertainty denotes the $1\sigma$ fitting error as shown in Fig.~\ref{fig:focusing}(c). This value is in good agreement with the prediction of the lensmaker's equation for a single refractive surface. For a plano-concave geometry in which light propagates through fused silica ($n = 1.46$ at 633~nm) and refracts at the concave silica-air interface, the focal length is given by $f = R/(n-1)$, yielding $f \approx \SI{231}{\micro\meter}$ for the measured ROC of \SI{106.13}{\micro\meter}. The $\sim$4\% discrepancy between the measured and calculated focal lengths is well within the error bar of the measurement.

\begin{figure*}
\includegraphics[width=\textwidth]{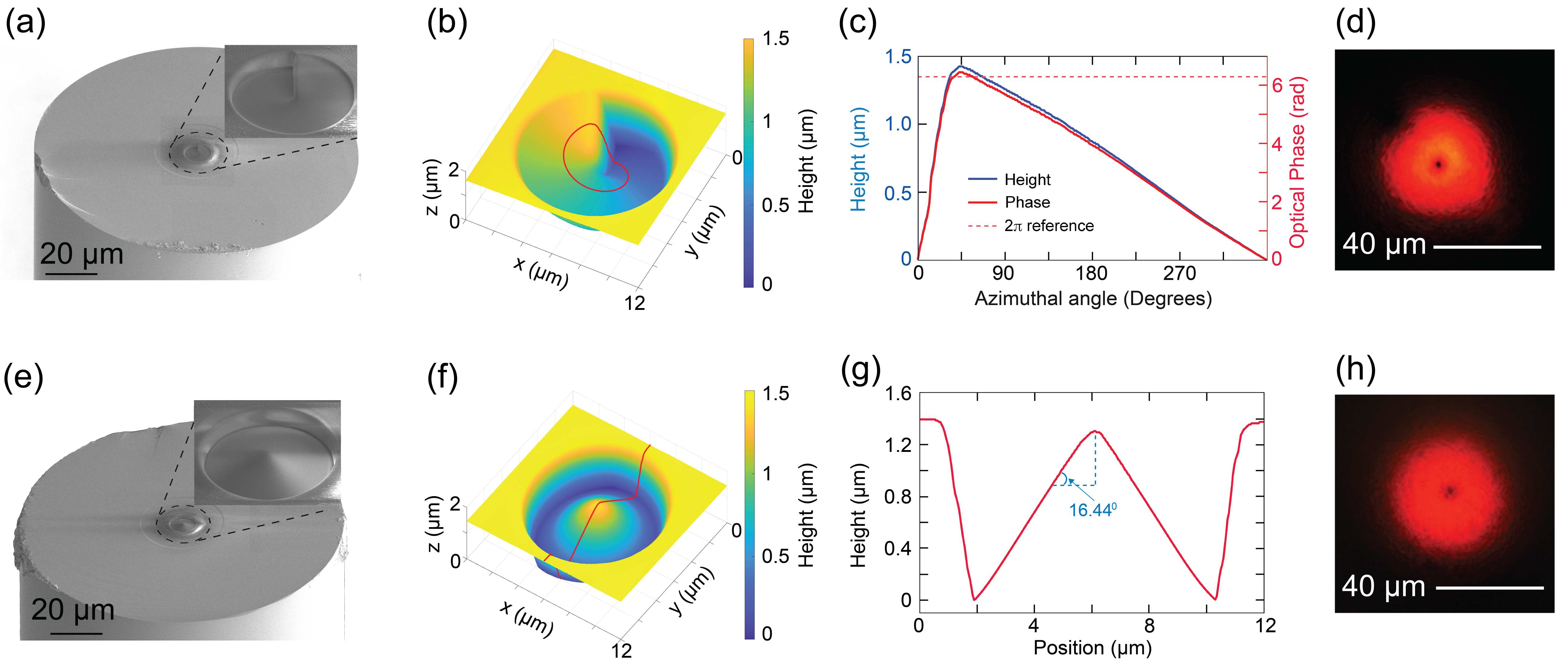}
\caption{Deterministic fabrication of non-spherical phase elements on a fiber facet. \textbf{(a,e)}~SEM images of the micro-spiral and micro-axicon elements with a diameter of \SI{10}{\micro\meter} fabricated on the fiber core, respectively. Inset: Magnified views. \textbf{(b,f)}~AFM topography images (scan area: $12~\si{\micro\meter} \times 12~\si{\micro\meter}$) for micro-spiral and micro-axicon features, respectively, with a red line across the center showing the location of line cuts used for subsequent analysis. \textbf{(c)}~Line cut height data across the azimuthal direction of the micro-spiral, calculated optical phase along the azimuthal direction, and $2\pi$ reference shown for comparison. \textbf{(g)}~Radial profile of the axicon yielding an effective cone angle of $16.4^\circ$ (Design: $16.0^\circ$), measured over the $\approx 4~\si{\micro\meter}$ mode-field diameter of the fiber. \textbf{(d,h)}~Far-field intensity images of micro-spiral and micro-axicon fibers taken using a Helium-Neon laser (633~nm) based microscope, respectively. Scale bar: \SI{40}{\micro\meter}.}
\label{fig:phase}
\end{figure*}

The demonstrated focal length of \SI{222}{\micro\meter} is directly relevant to fiber-based cavity quantum electrodynamics, enabling the realization of near-concentric ($L = 2R \approx \SI{200}{\micro\meter}$) configurations. The near-concentric cavity provides the smallest cavity mode volumes, and thus a high atom-photon coupling strength for single-atom cavity QED experiments~\cite{Grinkemeyer2025}.

\begin{figure*}
\includegraphics[width=0.66\textwidth]{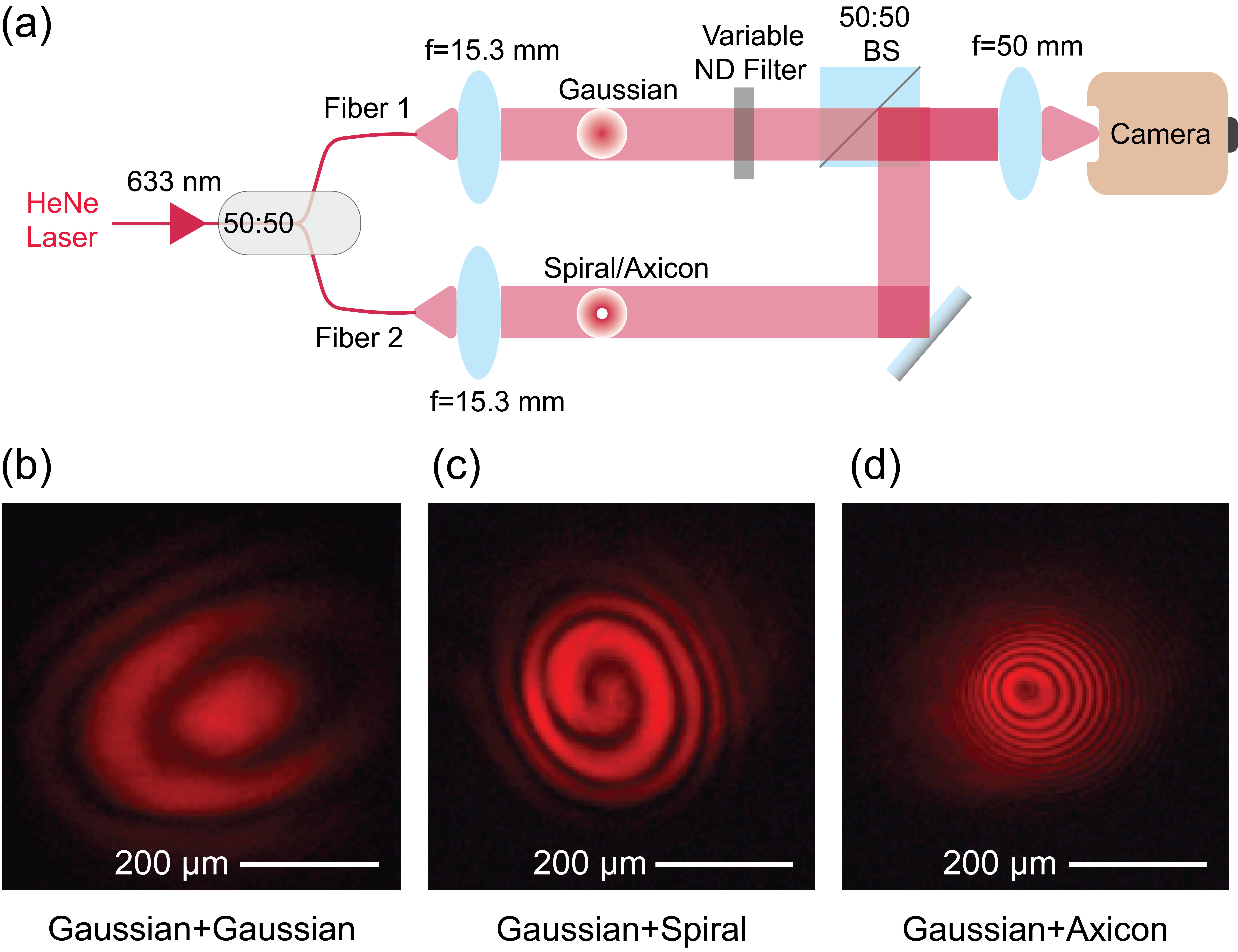}
\caption{Interferometric validation of phase modulation imposed by the fabricated elements. \textbf{(a)}~Mach-Zehnder interferometer used to interfere a reference Gaussian beam with the light transmitted through the structured fiber using a Helium-Neon laser (633 nm). The laser was split into two arms with a 50:50 1×2 fiber splitter (Thorlabs TW630R5F1). One arm was coupled to a standard flat-cleaved fiber patch cord to provide the reference Gaussian beam, while the other was coupled to the spiral or axicon fiber. \textbf{(b-d)}~Interference patterns of two Gaussian beams, one Gaussian plus one micro-spiral demonstrating presence of an azimuthal phase, and one Gaussian plus one micro-axicon pattern consistent with radial phase, respectively. Scale bar: \SI{200}{\micro\meter}.}
\label{fig:mzi}
\end{figure*}

\begin{figure*}[t]
\centering
\includegraphics[width=0.6\textwidth]{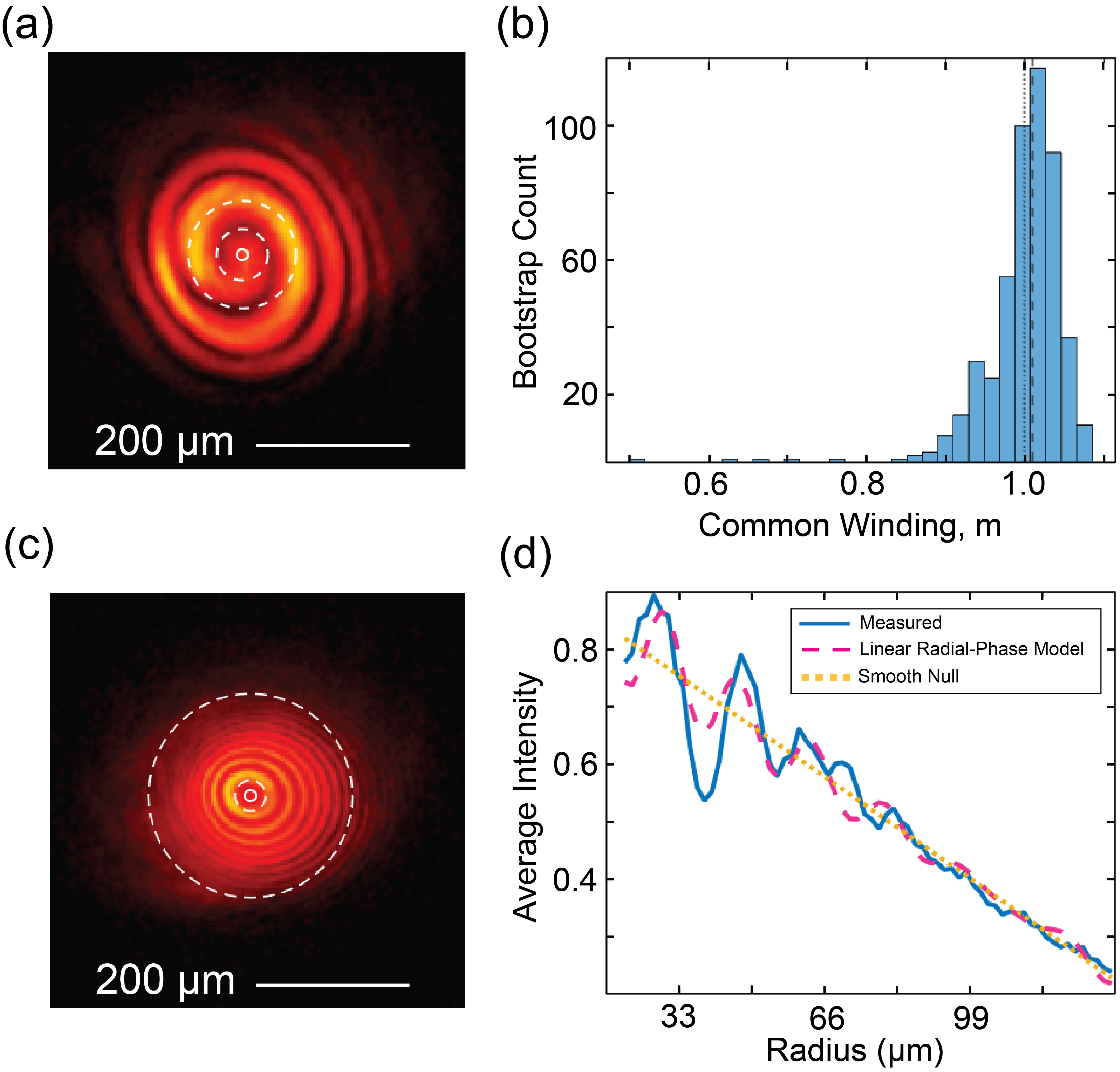}
\caption{Quantitative interferometric characterization of the FIB-fabricated micro-spiral and micro-axicon phase elements. \textbf{(a)}~Measured micro-spiral Mach-Zehnder interferogram with the selected finite-thickness annular analysis region indicated. The annulus spans radii of 10-21 pixels and contains 12 circular intensity sampling traces used for the common-winding analysis. \textbf{(b)}~Bootstrap distribution of the fitted common azimuthal winding parameter $m$, obtained by resampling the 12 circular traces within the selected annulus. The fit yields $m=1.0095$, with a 95\% bootstrap confidence interval of $0.8916$-$1.0600$. The vertical reference line indicates the ideal single-winding value $m=1$. \textbf{(c)}~Measured micro-axicon Mach-Zehnder interferogram with the selected annular analysis region indicated. The region excludes the central intensity maximum and low-signal outer portion while retaining the clearly resolved concentric interference fringes. \textbf{(d)}~Azimuthally averaged radial intensity profile within the selected region together with the fitted linear radial-phase model. Six radial intensity maxima are resolved, with ring positions varying linearly with fringe order ($R^2=0.99$). Here, $R^2$ is the coefficient of determination and quantifies the goodness of fit to the linear model, yielding a fitted radial fringe spacing of 5.07 pixels and an effective radial phase gradient of $1.24~\mathrm{rad/pixel}$. The fitted profile reproduces the measured radial intensity with $R^2=0.95$ and NRMSE $=0.06$. Here, NRMSE denotes the Normalized Root Mean Square Error of the fit.}
\label{fig:phase_quant}
\end{figure*}

\subsection{Micro-spiral phase plate}

The nanoscale, free-form fabrication capability of the FIB process is exploited to realize optical elements that directly tailor the phase of the Gaussian mode emerging from the fiber facet. As a first example, a micro-spiral phase plate is designed to impart an azimuthal phase of $2\pi$ (topological charge $\ell = 1$) on the transmitted beam at a wavelength of 633 nm, converting the fundamental Gaussian mode into a donut-shaped vortex beam carrying orbital angular momentum (OAM). The SEM image in Fig.~\ref{fig:phase}(a) shows the fabricated element with a diameter of \SI{10}{\micro\meter} on the fiber core, with the characteristic helical ramp clearly visible in the inset magnified view. AFM measurement in Fig.~\ref{fig:phase}(b) confirms smooth topography and height variation across the element, and the extracted azimuthal line cut [Fig.~\ref{fig:phase}(c)] shows excellent agreement between the measured optical phase (calculated from the height profile using the known refractive index of 1.46) and the ideal $2\pi$ reference. Far-field imaging using a He-Ne laser (633~nm) of the micro-spiral fiber output as shown in Fig.~\ref{fig:phase}(d) reveals the expected donut intensity profile characteristic of an OAM beam, confirming the successful generation of a vortex beam directly from the fiber facet. The ability to make spiral phase elements on the facets of single-mode fibers has potential applications in creating hybrid cavities to explore non-trivial optical response of cold atom systems~\cite{Hao2025}. Further, the direct generation of topologically robust donut beams has applications in free-space quantum links ~\cite{McLaren2014,Wang2022},  as well as alignment and characterization of state-of-the-art optical systems \cite{Song1999}.

\subsection{Micro-axicon}

The micro-axicon is designed to produce a Bessel-like beam, which is of interest for applications in optical trapping~\cite{Manek1998,Mukherjee2017}. The SEM image in Fig.~\ref{fig:phase}(e) and AFM measurement in Fig.~\ref{fig:phase}(f) confirm smooth topography and conical profile of the fabricated element. The extracted height line cut [Fig.~\ref{fig:phase}(g)] yields a cone half-angle of $16.44^\circ$ at a position spanning the \SI{4}{\micro\meter} mode-field diameter of the fiber. The far-field image using a He-Ne laser (633~nm) based microscope as shown in Fig.~\ref{fig:phase}(h) exhibits the concentric ring pattern characteristic of a Bessel beam, consistent with the expected behavior of an axicon-shaped element. 

\subsection{Interferometric Phase Characterization}

To directly verify the phase profiles imparted by the micro-spiral and micro-axicon elements, we construct a fiber-based Mach-Zehnder interferometer using a Helium-Neon (633 nm) laser as shown in Fig.~\ref{fig:mzi}(a). The laser beam is divided into two paths by a 50:50 $1 \times 2$ fiber splitter (Thorlabs TW630R5F1). The reference arm was routed through a standard flat-cleaved fiber patch cord to produce a Gaussian output beam, whereas the signal arm was routed through the fabricated spiral or axicon fiber. FC/PC fiber-mating sleeves (Thorlabs ADAFC2) were used to connect the splitter outputs to the respective fiber components, providing robust alignment and repeatable optical coupling. The two beams are collimated, combined on a 50:50 non-polarizing beamsplitter (Thorlabs BS004), and imaged onto a camera (Thorlabs CS165CU/M). The reference interference pattern of two Gaussian beams [Fig.~\ref{fig:mzi}(b)] exhibits the expected circular fringes. When the micro-spiral fiber is placed in one arm [Fig.~\ref{fig:mzi}(c)], the coaxial interference pattern displays a spiral profile~\cite{Ribeiro2016} with exactly one fringe per azimuthal rotation of $2\pi$, providing direct evidence of the helical phase front associated with an OAM state of topological charge $\ell = 1$. The interference of a Gaussian beam with the micro-axicon output as shown in Fig.~\ref{fig:mzi}(d) produces a pattern of concentric circular fringes, consistent with the radial phase imparted by the micro-axicon element.

\subsection{Quantitative phase characterization}

To quantitatively characterize the phase modulation produced by micro-spiral and micro-axicon elements, we analyze the measured Mach-Zehnder interferograms of both structures, as summarized in Fig.~\ref{fig:phase_quant}. For the micro-spiral, the interferogram is analyzed over a finite-thickness annular region centered on the phase singularity [Fig.~\ref{fig:phase_quant}(a)]. The selected annulus extends from an inner radius of 10 pixels to an outer radius of 21 pixels and contains 12 circular intensity traces. For each sampled radius, the azimuthal intensity is described by $I_j(\phi)=A_j+B_j\cos(m\phi+\phi_j)$, where $A_j$, $B_j$, and $\phi_j$ account for radius-dependent background intensity, fringe amplitude, and phase offset, respectively, while the winding parameter $m$ is common to all traces. An ideal single-winding spiral corresponds to $m=1$. A simultaneous fit to the 12 traces yields $m=1.0095$, corresponding to an accumulated phase of approximately $2.019\pi$ i.e.~a deviation of less than $1\%$ from the ideal single-winding value. To estimate the uncertainty associated with radial variation within the selected annulus, the circular traces from this fixed region of interest are resampled with replacement, meaning that each bootstrap sample is formed by randomly selecting traces from the original set while allowing individual traces to be selected more than once. The common-winding fit is then recalculated for each bootstrap sample. The resulting distribution [Fig.~\ref{fig:phase_quant}(b)] gives a 95\% bootstrap confidence interval of $0.8916 \leq m \leq 1.0600$. Since the ideal value $m=1$ lies within this interval, the measured interferogram is statistically consistent with one complete azimuthal phase winding. For the micro-axicon, an annular region centered on the concentric interference pattern is selected [Fig.~\ref{fig:phase_quant}(c)] such that the central intensity maximum and low-signal outer region are excluded while the clearly resolved concentric fringes are retained. Azimuthal averaging of this region produces the radial intensity profile shown in Fig.~\ref{fig:phase_quant}(d). The radial profile is fitted using a model corresponding to an approximately linear radial phase. Six radial intensity maxima are resolved, and their positions vary linearly with fringe order with $R^2=0.99$, yielding a fitted radial fringe spacing of 5.07 pixels and an effective radial phase gradient of $1.24~\mathrm{rad/pixel}$. The fitted radial-phase model reproduces the measured profile with $R^2=0.95$ and a normalized root-mean-square error of 0.06.

\subsection{Insertion Loss}

\begingroup

To quantify the optical loss introduced specifically by the FIB-fabricated micro-optical elements, we perform a measurement on the same fiber assembly before and after fabrication. Each device is fabricated from a single-mode fiber patch cord terminated by an FC/PC connector at the input and a freshly cleaved fiber facet at the output. A He-Ne laser ($\lambda=633$~nm) is coupled to the FC/PC input, and the optical power emitted from the pristine cleaved facet is first measured using an integrated sphere power sensor (Thorlabs S142C), denoted by $P_{ \mathrm{ref}}$ below. The micro-optical element is subsequently fabricated directly on this same output facet, without modifying the remainder of the fiber assembly, and the transmitted power is measured again under the same conditions and denoted $P_{ \mathrm{FIB}}$. Using the same FC/PC-terminated fiber assembly before and after FIB fabrication isolates the excess loss introduced by the modified facet. The excess insertion loss associated with fabrication of the micro-optical element is calculated as

\begin{equation}
    \mathrm{IL}
    =-10\log_{10}
    \left(
    \frac{P_{ \mathrm{FIB}}}{P_{\mathrm{ref}}}
    \right).
    \label{eq:insertion_loss}
\end{equation}

\begin{table*}[t]
\centering
\caption{Measured excess insertion loss of FIB-fabricated micro-optical
elements at $\lambda=633$~nm. $P_{\mathrm{ref}}$ is the optical power
measured from the pristine cleaved fiber facet before FIB fabrication, and
$P_{\mathrm{FIB}}$ is the power measured from the same fiber after
fabrication of the micro-optical element. The insertion loss is calculated
using Eq.~(\ref{eq:insertion_loss}). The uncertainties quoted for the
averaged insertion losses correspond to the standard deviation among the
measured devices.}
\label{tab:insertionloss}
\begin{tabular}{lcccc}
\toprule
Element &
$P_{\mathrm{ref}}$ (mW) &
$P_{\mathrm{FIB}}$ (mW) &
Transmission (\%) &
IL (dB) \\
\midrule
Spiral 1 & 1.711 & 1.571 & 91.82 & 0.37 \\
Spiral 2 & 2.243 & 1.961 & 87.42 & 0.58 \\
Spiral 3 & 1.428 & 1.327 & 92.92 & 0.32 \\
\midrule
Axicon 1 & 1.880 & 1.740 & 92.55 & 0.34 \\
Axicon 2 & 2.400 & 2.280 & 95.00 & 0.22 \\
Axicon 3 & 1.510 & 1.390 & 92.05 & 0.36 \\
Axicon 4 & 0.237 & 0.204 & 86.07 & 0.65 \\
\midrule
\multicolumn{3}{l}{Spiral average} &
90.72 & $0.42\pm0.14$ \\
\multicolumn{3}{l}{Axicon average} &
91.42 & $0.39\pm0.18$ \\
\multicolumn{3}{l}{All structures} &
91.12 & $0.41\pm0.15$ \\
\bottomrule
\end{tabular}
\end{table*}

As summarized in Table~\ref{tab:insertionloss}, the three micro-spiral devices exhibit transmission efficiencies between $87.4\%$ and $92.9\%$, corresponding to an average excess insertion loss of
$0.42\pm0.14$~dB. Similarly, the four micro-axicon devices exhibit transmission efficiencies between $86.1\%$ and $95.0\%$, with an average excess insertion loss of $0.39\pm0.18$~dB. Considering all seven measurements, the structures retain an average of approximately $91.1\%$ of the power transmitted by the pristine cleaved fiber, corresponding to a mean excess insertion loss of only $0.41\pm0.15$~dB. These measurements demonstrate that the excess insertion loss for our micro-machined structures is on the the sub-dB level.

The measured losses compare favorably with other fiber-facet micro-optical approaches. For example, Maier \textit{et al.} demonstrated non-monolithic facet-attached micro-optical elements fabricated by multiphoton 3D lithography and reported coupling losses as low as approximately $0.35$~dB \cite{Maier2022FacetOptics}. Although the device geometries and definitions of optical loss are not directly equivalent, both approaches achieve similar sub-dB loss regime. The underlying integration strategies, however, are fundamentally different. Facet-attached 3D-printed micro-optics employ an additive heterogeneous approach in which a polymer photoresist structure is formed on top of the existing fiber facet. In contrast, the present FIB process directly sculpts the optical functionality into the native silica fiber itself. The resulting structure is therefore monolithic and material-native, requiring neither an attached optical body nor an additional polymer-silica material interface. This geometry reduces the overall device footprint, removes the alignment degrees of freedom associated with a separate optical element, and avoids considerations such as adhesion and differential thermal or mechanical properties between the fiber and an added micro-optic.

Similarly, facet-attached 3D-printed microlens systems have reported an estimated $\sim0.4$~dB Fresnel contribution from two uncoated microlens surfaces alone \cite{Xu2023FacetMicrolens}. The present approach instead incorporates the desired optical functionality into the already existing fiber output surface. Thus, the measured $\sim0.4$~dB excess loss provides complex beam shaping with an optical-loss budget comparable to the interface loss alone of an additional uncoated micro- or bulk-optical element.

The distinction between additive polymer micro-optics and direct structuring of silica is particularly relevant for fiber-integrated experiments operating under ultra-high-vacuum (UHV) conditions, including neutral atom, trapped-ion, and cavity-QED platforms. Fused silica is an inorganic material routinely employed in vacuum optical assemblies, whereas polymeric materials require material-specific qualification of outgassing, thermal stability, and bake-out compatibility. Studies of 3D-printed polymers for UHV applications have consequently emphasized the need for explicit outgassing and residual-gas characterization before their use in such environments \cite{Domingues2025UHV}. The multiphoton-lithography approaches discussed above, for example, rely on polymerized photoresists to form the facet-attached optical structure \cite{Maier2022FacetOptics,Xu2023FacetMicrolens}. In contrast, FIB fabrication introduces no additional polymeric optical body at the fiber facet, providing a clear materials advantage for compact in-vacuum fiber-optic architectures.

\endgroup

\section{Conclusion}

\begin{figure*}
\includegraphics[width=0.66 \textwidth]{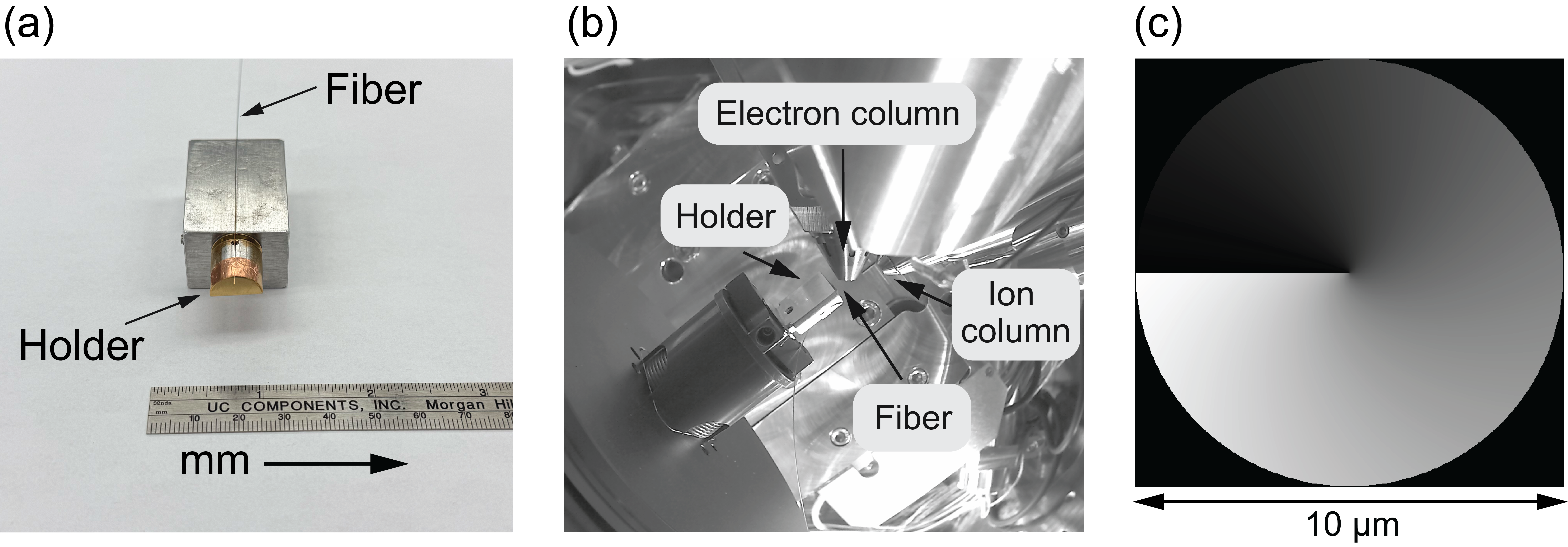}
\caption{Implementation of fiber-compatible FIB processing. \textbf{(a)}~Custom mount for holding cleaved fibers during FIB processing. Ruler for scale. \textbf{(b)}~Installed configuration inside a Helios G5 dual-beam system showing relative orientation of electron and ion columns. \textbf{(c)}~A typical design BMP file ($512 \times 512$ pixels) for a micro-spiral element with a diameter of \SI{10}{\micro\meter}.}
\label{fig:appendixA}
\end{figure*}

We present a single-step, versatile, and high-precision focused ion beam (FIB) micro-machining technique for fabricating micro-optical elements directly on the core of single-mode optical fibers. Central to this method is the precise ion-beam alignment achieved through eucentric height optimization at the fiber facet, together with iterative nano-scale AFM metrology used to precisely refine the target surface profiles. Unlike previous demonstrations that largely relied on fabrication alone with limited nano-scale feedback, our approach couples controlled beam alignment with repeated high-resolution metrology to achieve reproducible and highly accurate structures on the fiber facets. Our comprehensive characterization of the resulting micro-optical elements demonstrates excellent structural and optical accuracy. 

Our surface roughness analysis indicates that the FIB process preserves nanoscale surface quality at wavelengths relevant to quantum optics, minimizing scattering losses.
We note that residual implanted Ga\textsuperscript{+} ions from the FIB process may contribute to charging or other surface effects, and measurements to assess their presence and impact are underway. Further, prior studies have demonstrated that Ga$^{+}$-FIB-machined silica fiber surfaces can subsequently support functional dielectric coatings~\cite{Micco2015,Yan2024}, which is important for the development of high-finesse optical micro-cavities. Future process refinements may include a brief post-fabrication etch to remove a thin surface layer (on the order of tens of nanometers) consistent with the expected implantation depth. This work establishes FIB-fabricated fiber micro-optics as a precise and flexible platform for quantum information science applications, including fiber microcavities, structured-light generation, and neutral atom trapping and manipulation.

\begin{figure*}
\includegraphics[width=0.66 \textwidth]{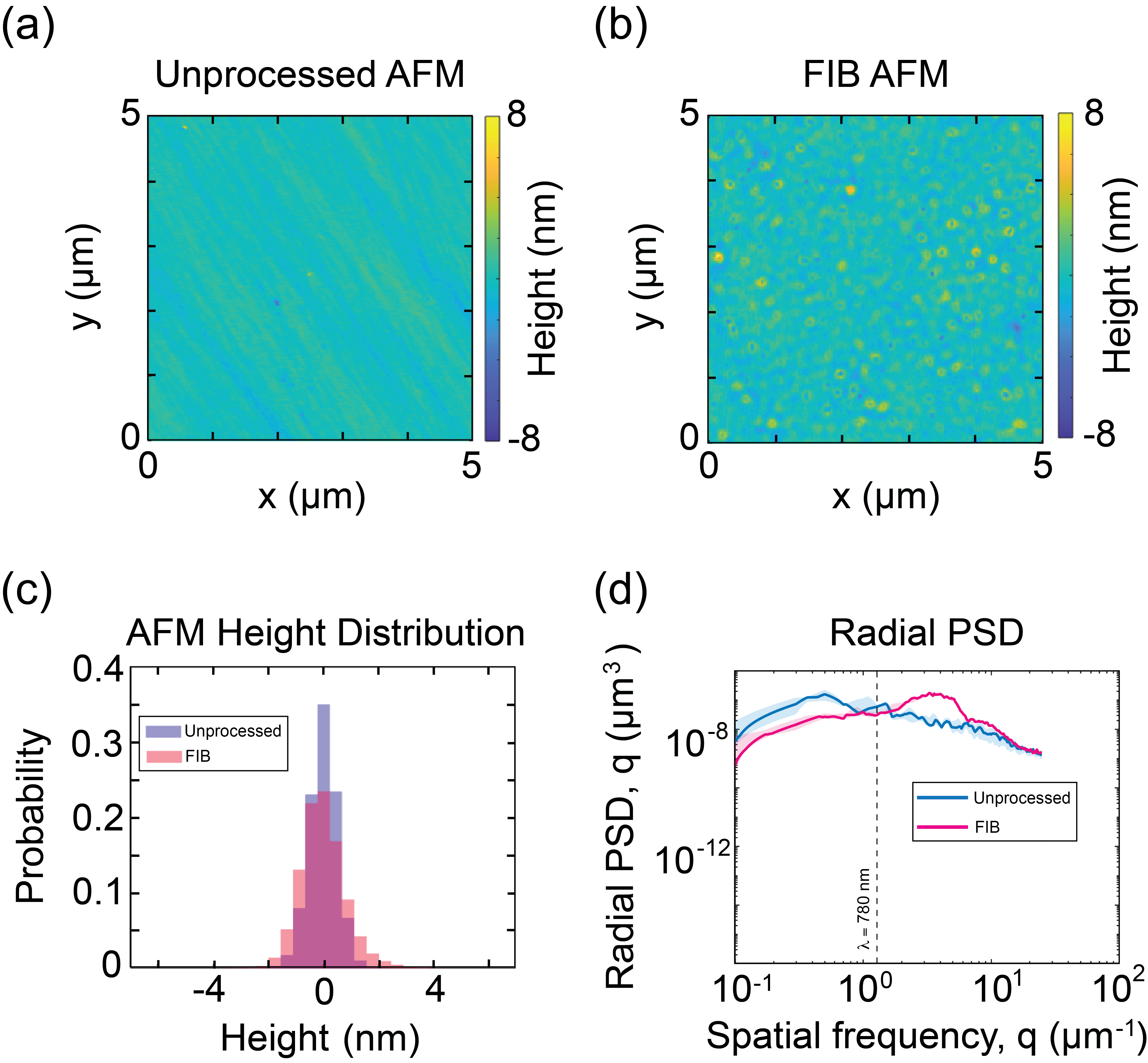}
\caption{Comparison of nanoscale roughness before and after FIB processing. \textbf{(a,b)}~AFM measurements (scan area: $5~\si{\micro\meter} \times 5~\si{\micro\meter}$) on an unprocessed and FIB-processed surface area for comparison of nanoscale surface roughness. \textbf{(c)}~Height distribution histograms showing comparable rms roughness. \textbf{(d)}~Radially averaged power spectral density (PSD) obtained from four unprocessed and five FIB-processed surface measurements. Solid curves show the median PSD for each group, while the shaded regions indicate the 25th-75th percentile range across the independent scans. The vertical dashed line at $q \approx 1.28~\si{\micro\meter}^{-1}$ corresponds to a lateral surface period of 780~nm and is included as a characteristic optical length-scale reference.}
\label{fig:roughness}
\end{figure*}
\begin{acknowledgments}
This work was supported by Brookhaven National Laboratory (BNL) under LDRD 24-055. The authors acknowledge access to the Helios G5 Dual Beam FIB system used for fabrication and SEM characterization at the Center for Functional Nanomaterials (CFN), which is a U.S. Department of Energy (DOE) Office of Science User Facility, at BNL under Contract No.~DE-SC0012704. The authors thank Hong Li (Interdisciplinary Science Department) for assistance with AFM measurements and William Weldon (Instrumentation Department) for support in the design and fabrication of the fiber holders.
\end{acknowledgments}

\section*{Author Declarations}
\subsection*{Conflicts of Interest}
The authors have no conflicts of interest to disclose.

\subsection*{Data Availability}
The data supporting the experiments and analysis in this paper are available from the corresponding author upon reasonable request.

\appendix
\section{Fiber Mounting and Pattern Design}
\label{app:mounting}

Figure~\ref{fig:appendixA}(a) shows the custom-designed aluminum fiber holder (approximately $1~\mathrm{in} \times 0.5~\mathrm{in}$) used to secure short sections of the fiber during processing; a mounted fiber segment is also visible. The holder, together with the fiber, is introduced into the Helios G5 dual-beam FIB chamber [Fig.~\ref{fig:appendixA}(b)], where the relative geometry of the electron and ion columns with respect to the fiber facet is indicated. The stage is tilted by $52^\circ$ to align the fiber facet normal to the incident ion beam, ensuring perpendicular milling under standard dual-beam operating conditions. A representative bitmap design used for fabricating a \SI{10}{\micro\meter} diameter micro-spiral element is shown in Fig.~\ref{fig:appendixA}(c), illustrating the grayscale encoding employed to translate the desired height profile into a spatially varying ion dose. After fabrication of micro-optical elements, the fiber section is spliced to a single-mode fiber patch cable for characterization using a Fujikura FSM-100P+ fusion splicer.

\section{Surface Roughness Analysis}
\label{app:roughness}

AFM height maps $z(x,y)$ were acquired over $5~\si{\micro\meter} \times 5~\si{\micro\meter}$ regions for both unprocessed and FIB-processed areas on silica substrate (Corning 7980) as shown in Figs.~\ref{fig:roughness}(a) and~\ref{fig:roughness}(b), respectively. Prior to roughness analysis, the topography was leveled by removing the best-fit plane (and optional line-by-line flattening) to suppress scan tilt and low-order drift, after which areal roughness metrics were computed. The areal RMS roughness $S_q$ was evaluated as the standard deviation of the leveled height distribution,
\begin{equation}
S_q = \sqrt{\langle (z - \langle z \rangle)^2 \rangle},
\label{eq:roughness}
\end{equation}
where $\langle \cdot \rangle$ denotes the spatial average across all pixels. Using Eq.~(\ref{eq:roughness}), we obtain { $S_q = \SI{0.414} {\nano\meter}$} for unprocessed and {$S_q = \SI{0.855} {\nano\meter}$} for FIB-processed area. The difference {$\Delta S_q = \SI{0.441} {\nano\meter}$} indicates an added sub-nanometer roughness which is negligible in comparison to optical wavelengths. This is also corroborated by height distribution histograms in Fig.~\ref{fig:roughness}(c), which show that the changes in the distribution are negligible. Notably, the additional roughness remains at a level comparable to that reported in earlier studies~\cite{Maier2025} which employed multi-step fabrication workflows, while our method attains similar surface quality through a single-step FIB micro-machining.

To probe whether FIB introduces roughness preferentially at specific lateral length scales (which would not be captured by a single scalar $S_q$), we computed the two-dimensional surface power spectral density (PSD) from the Fourier transform of the leveled topography~\cite{Schroder2011},
\begin{equation}
\mathrm{PSD}(\bm{q}) = \frac{1}{A}\left|\mathcal{F}\{z(x,y)\}\right|^2, \quad \bm{q} = (q_x, q_y),
\label{eq:psd}
\end{equation}
and for approximately isotropic surfaces we report the azimuthally averaged (``radial'') PSD, $\mathrm{PSD}(q) = \langle \mathrm{PSD}(\bm{q})\rangle_{|\bm{q}|=q}$, which compactly describes roughness power versus spatial frequency~\cite{Schroder2011}. In our measurements, the unprocessed and FIB radial PSDs overlap closely as shown in Fig.~\ref{fig:roughness}(d) across the accessible spatial-frequency band, indicating that FIB does not introduce new roughness components at any characteristic lateral scale relevant to optical wavelengths.


To evaluate the reproducibility of the roughness measurements and quantify the surface modification introduced by FIB processing, we repeated the AFM analysis over multiple independently measured regions. Four $5~\si{\micro\meter}\times5~\si{\micro\meter}$ regions were measured on unprocessed silica, while five independent regions were measured after FIB processing. For each scan, a second-order two-dimensional polynomial was removed to suppress macroscopic sample/scanner curvature, followed by a linear line-by-line correction along the experimentally defined fast-scan direction. No spatial smoothing or selective removal of surface features was applied before calculating the roughness metrics.

The individual areal RMS roughness values, $S_q$, are summarized in Table~\ref{tab:roughness_statistics}. The unprocessed surfaces exhibit $S_q$ values between $0.41$ and $0.67$~nm, with an average of

\begin{equation}
S_{q,\mathrm{unprocessed}}
=
0.56 \pm 0.11~\mathrm{nm},
\end{equation}

where the uncertainty denotes the standard deviation across the four independent measurements. Following FIB processing, the five measured regions exhibit $S_q$ values between $0.80$ and $0.86$~nm, yielding

\begin{equation}
S_{q,\mathrm{FIB}}
=
0.83 \pm 0.03~\mathrm{nm}.
\end{equation}

The corresponding absolute increase in RMS roughness is therefore only

\begin{equation}
\Delta S_q
=
S_{q,\mathrm{FIB}}
-
S_{q,\mathrm{unprocessed}}
\approx
0.27~\mathrm{nm}.
\end{equation}

Although the repeated measurements resolve a systematic increase in roughness following ion-beam processing, the magnitude of the added roughness remains sub-nanometer and approximately three orders of magnitude smaller than the optical wavelengths of interest. The comparatively small standard deviation of the FIB-processed measurements ($\sim0.03$~nm) further indicates that the resulting surface roughness is highly reproducible across independently measured regions.

\begin{table}[t]
\centering
\caption{Areal RMS roughness, $S_q$, obtained from independent
$5~\si{\micro\meter}\times5~\si{\micro\meter}$ AFM measurements of
unprocessed and FIB-processed silica surfaces. The reported group
uncertainties correspond to the standard deviation among independent
measurements.}
\label{tab:roughness_statistics}
\begin{tabular}{lcc}
\toprule
Surface condition & Scan & $S_q$ (nm) \\
\midrule
Unprocessed & Sample 1 & 0.414 \\
            & Sample 2 & 0.671 \\
            & Sample 3 & 0.583 \\
            & Sample 4 & 0.583 \\
\cmidrule(lr){2-3}
            & Mean $\pm$ SD & $0.563 \pm 0.107$ \\
\midrule
FIB processed & FIB 1 & 0.855 \\
              & FIB 2 & 0.855 \\
              & FIB 3 & 0.814 \\
              & FIB 4 & 0.797 \\
              & FIB 5 & 0.832 \\
\cmidrule(lr){2-3}
              & Mean $\pm$ SD & $0.831 \pm 0.025$ \\
\bottomrule
\end{tabular}
\end{table}
To probe whether FIB introduces roughness preferentially at specific lateral length scales (which would not be captured by a single scalar $S_q$), we computed the two-dimensional surface power spectral density (PSD) from the Fourier transform of the leveled topography~\cite{Schroder2011},
\begin{equation}
\mathrm{PSD}(\bm{q}) = \frac{1}{A}\left|\mathcal{F}\{z(x,y)\}\right|^2, \quad \bm{q} = (q_x, q_y),
\label{eq:psd}
\end{equation}
and for approximately isotropic surfaces we report the azimuthally averaged (``radial'') PSD, $\mathrm{PSD}(q) = \langle \mathrm{PSD}(\bm{q})\rangle_{|\bm{q}|=q}$, which compactly describes roughness power versus spatial frequency~\cite{Schroder2011}. In Fig.~\ref{fig:roughness}(d), the solid curves represent the median radial PSD obtained from the four unprocessed and five FIB-processed measurements, respectively, while the shaded regions indicate the 25th-75th percentile range and therefore visualize the scan-to-scan variation. The vertical reference line at $q=1/0.780~\si{\micro\meter}^{-1}\approx1.28~\si{\micro\meter}^{-1}$ corresponds to a lateral surface period of 780~nm and is included as a characteristic optical length-scale reference. Taken together with the RMS roughness statistics, the PSD measurements show that FIB processing does not produce a large broadband degradation of the silica surface, but rather introduces only an extremely small increase in the integrated nanoscale roughness.

\bibliography{references}

\end{document}